\documentclass[symmetry,article,accept,moreauthors,pdftex]{mdpi} 

\firstpage{1} 
\pubvolume{1}
\issuenum{1}
\articlenumber{0}
\pubyear{2021}
\copyrightyear{2020}
\datereceived{} 
\dateaccepted{} 
\datepublished{} 
\hreflink{https://doi.org/} % If needed use \linebreak
\usepackage{mathtools}
\usepackage{cancel}

\Title{Operator relationship between conventional coupled cluster and unitary coupled cluster}

\TitleCitation{Operator relationship between conventional coupled cluster and unitary coupled cluster}

\Author{James K. Freericks $^{1}$\orcidA{} %Firstname Lastname $^{1,\ddagger}$ and Firstname Lastname $^{2,}$*
}

\AuthorNames{James K. Freericks}

\AuthorCitation{Freericks, J. K.}
\address{%
$^{1}$ \quad Department of Physics, Georgetown University, 37th and O Sts. NW, Washington, DC 20057 USA; james.freericks@georgetown.edu}

\corres{Correspondence: james.freericks@georgetown.edu
}

\abstract{The chemistry community has long sought the exact relationship between the conventional and the unitary coupled cluster ansatz for a single-reference system, especially given the interest in performing quantum chemistry on quantum computers. In this work, we show how one can use the operator manipulations given by the exponential disentangling identity and the Hadamard lemma to relate the factorized form of the unitary coupled-cluster approximation to a factorized form of the conventional coupled cluster approximation (the factorized form is required, because some amplitudes are operator-valued and do not commute with other terms). By employing the Trotter product formula, one can then relate the factorized form to the standard form of the unitary coupled cluster ansatz. The operator dependence of the factorized form of the coupled cluster approximation can also be removed at the expense of requiring even more higher-rank operators, finally yielding the conventional coupled cluster. The algebraic manipulations of this approach are daunting to carry out by hand, but can be automated on a computer for small enough systems.}


\begin{document}
%%%%%%%%%%%%%%%%%%%%%%%%%%%%%%%%%%%%%%%%%%
%\setcounter{section}{-1} %% Remove this when starting to work on the template.
\section{Introduction}

The coupled cluster (CC) approach~\cite{bartlett_review} is regarded as the gold standard for quantum chemistry, especially as it is applied to weakly correlated molecular systems.   In this work, we will focus entirely on a single-reference CC ansatz, where a series of operators are applied to a reference state, which is a product state of $N_e$ occupied spin-orbitals, such as given by a Hartree-Fock calculation.  There are a number of innovative keys to the CC approximation in quantum chemistry. First, it provides a low-rank representation of a many-body quantum state that is size-consistent for closed-shell fragments (unlike many configuration interaction approximations) meaning it reduces to the closed shell atomic systems when the molecule is pulled apart by stretching. Second, it is size extensive, implying it has a linked-cluster-like expansion in terms of diagrams, so it scales the energy properly in the thermodynamic limit. Third, it is extremely efficient in its computational algorithm, because it never works with the explicit wavefunction. It instead uses a similarity transformation of the Hamiltonian, calculated via the Hadamard lemma (also called the Baker-Campbell-Hausdorff formula), which truncates at the fourth order term in the expansion (because the Hamiltonian has at most four fermionic operators in each term). This allows the algorithm to work solely with the CC amplitudes, rather than with the wavefunction. As a result, implementation of this approach, while complicated, is extremely efficient. One drawback of the CC approach is that it is not a variational calculation and it often fails when correlations become too strong. The unitary variant of the coupled cluster approximation (called unitary coupled cluster, abbreviated to UCC) is also a low-rank representation of the wavefunction that is size consistent and size extensive, but because one must perform a variational calculation using the full wavefunction, calculations with UCC are significantly (exponentially) less efficient than their CC counterparts on conventional computers. Yet, because quantum computers work most efficiently with unitary operations, the UCC ansatz is the only approach that can be practically implemented on a quantum computer. 

Ever since the UCC was introduced~\cite{kutzelnig,bartlett}, people have wondered what is the exact relationship between it and the conventional CC approximation? While no precise relationship has been found, Paldus and collaborators performed an interesting analysis based on group theory representations~\cite{paldus} and work on model Hamiltonians explicitly showed that the two ansatzes are definitely different~\cite{scuseria}. In the mean time, quantum computing algorithms were found to implement the UCC in a factorized form~\cite{mazziotti,vqe,adapt,evangalista}, sometimes called a product form of the UCC, or the single-step of a Trotter product formula for the ansatz (more details will be given below). In this factorized form, which is similar to the form used to solve the anti-Hermitan Schr\"odinger equation~\cite{mazziotti}, one can show that the excitation and de-excitation operator, of arbitrary rank, have a ``hidden'' SU(2) symmetry~\cite{evangalista,xu,jia}, which has been employed to find an exact operator analogue of the Euler formula for complex exponentials. This hidden SU(2) symmetry also allows one to use the so-called exponential disentangling identities~\cite{atom}, which we will re-derive below in the context of the factorized form of UCC. This disentangling identity separates the excitation part of the UCC term from the de-excitation part. Once this has been accomplished for each UCC factor, we simply need to re-order the exponential terms, by using the Hadamard lemma, to move all excitation operators to the left; the corresponding de-excitation operators will annihilate against the reference state. In the process, we will also generate mixed terms, that no longer separate into pure excitations and de-excitations, but create terms that mix the two. Many of the mixed terms also annihilate against the reference state. This then produces the equivalent CC approximation---but with a specific difference---some of the amplitudes are operator-valued, implying they have different coefficients for different determinants, so are not in the traditional CC form. Another key observation is that a low-rank UCC ansatz in factorized form, will not typically map to a low-rank CC approximation---this shows that while there is an equivalence between the factorized form of the two ansatzes, restricting to low-rank approximations for one will produce different ansatzes for the other. In addition, some of the amplitudes in the mapping to the CC form have factors in them that are have functions of number operators as exponents---these terms show the two expressions are not completely equivalent, unless only one exponent of that operator expression is present in the CC ansatz; we have a counterexample, showing that generically, CC amplitudes \textit{will be} operator-valued in the equivalent formula. To relate this ansatz to the traditional UCC, requires simply applying the Trotter product formula and taking the limit as the number of steps goes to infinity; but because of the operator-valued CC amplitudes, we require further steps to relate the results to the traditional CC ansatz. We choose the appropriate exponent for the operator-valued term when it acts on the reference, and we use higher-rank CC terms to correct remaining terms that are represented improperly. At that point, we have related the traditional UCC to the traditional CC. 

There is significant work to carry out this approach to determine the operator equivalence and it rapidly becomes impossible to work results out analytically. Because of the complexity of the methodology discussed here, carrying out these approaches requires significant algebraic manipulations, which will be best handled by using computer-based algebraic manipulations. Implementing such a scheme is beyond this work. We will also describe a concrete algorithm, similar to the so-called elimination algorithm~\cite{evangalista}, which will allow one to directly relate the two ansatzes as well, but it does so by constructing the calculated UCC wavefunction using the conventional CC approach. 

%%%%%%%%%%%%%%%%%%%%%%%%%%%%%%%%%%%%%%%%%%
\section{Formalism and analysis of the hidden symmetry}

In this section, we will provide technical descriptions of the CC, UCC, and factorized form of the UCC approximations. We will then explore the hidden symmetry and derive the operator identities that emerge from this symmetry. We will then use these identities to create a fully disentangled operator identity, which then will be re-ordered using the Hadamard lemma.

We employ a second-quantized formalism to work with these systems. We typically use the spin orbitals of a Hartree-Fock (HF) approximation (although this is not a requirement) as the basis for the second-quantized operators. The creation (and annihilation) operators are
denoted $\hat{a}_i^\dagger$ ($\hat{a}_i^{\phantom{\dagger}}$) for the $N_e$ filled (or real) spin-orbitals and $\hat{a}_a^\dagger$ ($\hat{a}_a^{\phantom{\dagger}}$) for the unoccupied (or virtual) spin-orbitals. The indices $i$, $j$, $k~\cdots$, chosen from the middle of the alphabet, denote the orbitals occupied in the reference state $|\Psi_0\rangle=\prod_{i=1}^{N_e}\hat{a}_i^\dagger|0\rangle$, where $|0\rangle$ is the vacuum state, annihilated by all annihilation
operators. The indices $a$, $b$, $c~\cdots$, chosen from the beginning of the alphabet, denote the unoccupied orbitals to be used in the calculation. For both the occupied, and unoccupied spin-orbitals, we choose a specific ordering scheme for the indices that refer to each orbital. We will not discuss how the spin-orbitals are chosen in this work. In chemical calculations the single reference is often an unrestricted or restricted Hartree-Fock state. The difference between these two plays no role in the formal developments of this work, so we do not discuss this issue further. All we require is the product-state form of the reference state.

A rank-$n$ CC excitation operator is of the form
\begin{equation}
    \underbrace{\hat{a}_a^\dagger\hat{a}_{b\phantom{j}}^\dagger\cdots}_{n~\text{terms (virtual)}}~\underbrace{\cdots \hat{a}_j^{\phantom{\dagger}}\hat{a}_i^{\phantom{\dagger}}}_{n~\text{terms (real)}},
\end{equation}
where we have the ordering $a<b<c<\cdots$ for
the unoccupied orbitals and 
$i<j<k<\cdots$ for the occupied orbitals; we list all occupied orbitals before all virtual orbitals. You can see that when this excitation operator acts on the reference
state $|\Psi_0\rangle$, it will remove $n$ electrons from the occupied orbitals and place them in $n$ of the previously unoccupied virtual orbitals. Because these particles are fermions, the specific ordering convention we use to label the different spin-orbitals sets the overall sign of this contribution. In CC theory, we group together all possible terms according to a given rank and sum them together to create the excitation of a specific rank (here, a rank-$n$ example)
\begin{equation}
    \hat{T}_n=\sum_{i<j<k\cdots~}^{\text{real}}\sum_{a<b<c\cdots}^{\text{virtual}}\theta_{ijk\cdots}^{abc\cdots}\underbrace{\hat{a}_a^\dagger\hat{a}_{b\phantom{j}}^\dagger\cdots}_{n~\text{terms}}~\underbrace{\cdots \hat{a}_j^{\phantom{\dagger}}\hat{a}_i^{\phantom{\dagger}}}_{n~\text{terms}},
\end{equation}
where the real numbers $\theta_{ijk\cdots}^{abc\cdots}$ are called the rank-$n$ amplitudes. The conventional coupled cluster approximation uses all excitation operators of small rank (singles, corresponding to $n=1$ and denoted S and doubles, corresponding to $n=2$ and denoted D) and sometimes supplements them with selected
orbitals of higher rank (usually no more than triples (T) and quads (Q) and often they are treated perturbatively). Hence, a CCSD excitation operator is $\hat{T}(SD)=\hat{T}_1+\hat{T}_2$. The low-rank representation of the CC wavefunction is then $|\Psi_{CC}\rangle=e^{\hat{T}(SD)}|\psi_0\rangle$; any exponential of a sum of only excitation operators is called a CC ansatz---the approach becomes useful when accurate quantum chemistry calculations require only low-rank amplitudes in the representation of the wavefunction.

The exponential of the excitation operator $e^{\hat{T}}$, is not a unitary operator, because $\hat{T}^\dagger\ne-\hat{T}$. But, we can form a unitary exponential of an excitation minus a de-excitation operator via $e^{\hat{T}-\hat{T}^\dagger}$, so that the unitary wavefunction ansatz is $|\Psi_{UCC}\rangle=e^{\hat{T}-\hat{T}^\dagger}|\Psi_0\rangle$. As with the CC approximation, one usually chooses the amplitudes via some low-rank procedure such as choosing S and D, or even restricting to only important S and D terms (which then needs a criterion to determine whether a term is important). 

In general, there is no simple way to work directly with the UCC ansatz as written. One can expand the exponential in a power series and continue including terms of higher powers in the summation, until the results no longer change~\cite{ucc_exp_sum}. Then, if the amplitudes are not too large that one needs to worry about loss of precision, one can evaluate the UCC wavefunction in this fashion. One can also approximately evaluate the similarity transformation of the Hamiltonian, but truncated to some number of terms. An alternative, most useful for quantum computing, is to adopt a Trotter product formula evaluation. The Trotter product formula is the identity
\begin{equation}
    e^{\hat{T}-\hat{T}^\dagger}=\lim_{M\to\infty}\left ( \prod_{n}\prod_{i<j<k\cdots~}^{\text{real}}\prod_{a<b<c\cdots}^{\text{virtual}}e^{\frac{1}{M}\theta_{ijk\cdots}^{abc\cdots}\Big(\underbrace{\hat{a}_a^\dagger\hat{a}_{b\phantom{j}}^\dagger\cdots}_{n~\text{terms}}~\underbrace{\cdots \hat{a}_j^{\phantom{\dagger}}\hat{a}_i^{\phantom{\dagger}}}_{n~\text{terms}}-\underbrace{\hat{a}_i^\dagger\hat{a}_{j\phantom{j}}^\dagger\cdots}_{n~\text{terms}}~\underbrace{\cdots \hat{a}_b^{\phantom{\dagger}}\hat{a}_a^{\phantom{\dagger}}}_{n~\text{terms}}\Big)} \right )^M.
\end{equation}
It is an exact operator identity, so the ordering of the factors inside the big parenthesis does not matter, but the limit $M\to\infty$ is often not feasible to take. Then one selects a specific value of $M$ for an approximation. Unfortunately, in this case, the ordering matters, as can be clearly seen in the extreme limit of $M=1$, where we obtain a different result if we switch the order of two exponential factors that do not commute. This $M=1$ approximation is also called the factorized form of the UCC. One may think the $M=1$ case must be a poor approximation, but in many cases the variational principle can absorb many of the Trotter errors by changing the values of the amplitudes, making the factorized form an accurate and valid approximation in its own right for a wavefunction ansatz~\cite{ucc3}. Indeed, this type of approximation has already been used in the anti-Hermitian Schr\"odinger equation approach~\cite{mazziotti}.

There is a huge benefit in using the Trotter product formula---every factor in the product has a hidden SU(2) symmetry associated with it, which we discuss in more detail now. Note that this ``effective'' spin symmetry has nothing to do with the spin of the electrons, it is an operator symmetry derived from the commutation relations of the excitation and de-excitation operators of each UCC factor, when written in factorized form.  We define the following ``pseudospin'' operators for any rank-$n$ UCC factor via
\begin{equation}
    \hat{S}_+=i\underbrace{\hat{a}_a^\dagger\hat{a}_{b\phantom{j}}^\dagger\cdots}_{n~\text{terms}}~\underbrace{\cdots \hat{a}_j^{\phantom{\dagger}}\hat{a}_i^{\phantom{\dagger}}}_{n~\text{terms}}~~\text{and}~~\hat{S}_-=\hat{S}_+^\dagger=-i\underbrace{\hat{a}_i^\dagger\hat{a}_{j}^\dagger\cdots}_{n~\text{terms}}~\underbrace{\cdots \hat{a}_{b\phantom{j}}^{\phantom{\dagger}}\hat{a}_a^{\phantom{\dagger}}}_{n~\text{terms}}.
\end{equation}
We then define $\hat{S}_z$ via $[\hat{S}_+,\hat{S}_-]=2\hat{S}_z$, so that
\begin{align}
    \hat{S}_z&=\frac{1}{2}(\hat{S}_+\hat{S}_--\hat{S}_-\hat{S}_+)\nonumber\\
    &=\frac{1}{2}\Big (\underbrace{\hat{n}_a\hat{n}_{b\phantom{j}}\cdots}_{n~\text{terms}}~\underbrace{\cdots(1-\hat{n}_j)(1-\hat{n}_i)}_{n~\text{terms}}-\underbrace{\hat{n}_i\hat{n}_j\cdots}_{n~\text{terms}}~\underbrace{\cdots(1-\hat{n}_{b\phantom{j}})(1-\hat{n}_a)}_{n~\text{terms}}\Big ).
\end{align}
Here the number operator is $\hat{n}=\hat{a}^\dagger\hat{a}$, where we suppressed the index in the definition for simplicity. One can then immediately show that $[\hat{S}_z,\hat{S}_\pm]=\pm\hat{S}_\pm$. This establishes the SU(2) symmetry for the operators that appear in the exponent of the UCC factors.

In fact, because the Pauli exclusion principle requires $\hat{a}^2=(\hat{a}^\dagger)^2=0$, we see the following identities as well:
\begin{equation}
    (\hat{S}_\pm)^2=0,~~\hat{S}_+\hat{S}_-\hat{S}_+=\hat{S}_+,~~\text{and}~~\hat{S}_-\hat{S}_+\hat{S}_-=\hat{S}_-.
    \label{eq:idents}
\end{equation}
Any UCC factor, can then be written as
\begin{equation}
    e^{-i\theta_{ijk\cdots}^{abc\cdots}(\hat{S}_++\hat{S}_-)}=\sum_{n=0}^\infty
    \frac{\left (-i\theta_{ijk\cdots}^{abc\cdots}\right )^n}{n!}(\hat{S}_++\hat{S}_-)^n,
\end{equation}
by simply expanding the exponential in an infinite power series (the series always converges for real $\theta$, because the operators have a finite dimensional representation, and hence have a bounded norm). The identities in Eq.~(\ref{eq:idents}) then tell us that $(\hat{S}_++\hat{S}_-)^2=(\hat{S}_+)^2+\hat{S}_+\hat{S}_-+\hat{S}_-\hat{S}_++(\hat{S}_-)^2$, which is equal to $\hat{S}_+\hat{S}_-+\hat{S}_-\hat{S}_+$. Hence, $(\hat{S}_++\hat{S}_-)^3=(\hat{S}_+\hat{S}_-+\hat{S}_-\hat{S}_+)(\hat{S}_++\hat{S}_+)$, which is $\hat{S}_++\hat{S}_-$. So, we immediately learn that
\begin{equation}
    (\hat{S}_++\hat{S}_-)^n=\begin{cases}\mathbb{I},~&\text{if }n=0\\\hat{S}_++\hat{S}_-,~&\text{if }n=\text{odd}\\
    \hat{S}_+\hat{S}_-+\hat{S}_-\hat{S}_+,~&\text{if }n=\text{even and positive.}\end{cases}
\end{equation}
This allows the sum to be performed, and we find that
\begin{equation}
    e^{-i\theta(\hat{S}_++\hat{S}_-)}=\hat{\mathbb{I}}-i\sin\theta (\hat{S}_++\hat{S}_-)+(\cos\theta-1)(\hat{S}_+\hat{S}_-+\hat{S}_-\hat{S}_+).
    \label{eq:euler}
\end{equation}
This is a well-known SU(2) identity~\cite{evangalista,xu,jia} that generalizes the Euler identity $e^{-i\theta}=\cos\theta-i\sin\theta$ to operators. It is different from the conventional Pauli matrix identity, because it involves the direct sum of spin-0 and spin-$\frac{1}{2}$ representations---this is because spin-0 corresponds to the case where both $S_+$ and $S_-$ annihilate the state, while spin-$\frac{1}{2}$ can be raised or lowered only  once. Note further, the term $\hat{S}_+\hat{S}_-+\hat{S}_-\hat{S}_+$ acts as the identity operator on the spin-$\frac{1}{2}$ states---it is not the $\hat{S}_z$ operator.

It turns out that there is a second identity, called the exponential disentangling identity, that disentangles the exponential factors in a different way. Because SU(2) is a Lie algebra, we can prove the identity by proving it for the Pauli spin matrices---then group theory tells us it holds for all representations, because the Pauli spin matrices are a faithful representation of SU(2). Recall that the Pauli matrices are
\begin{equation}
    \sigma_+=\begin{pmatrix}0&2\\0&0\end{pmatrix},~~\sigma_-=\begin{pmatrix*}[r]0&0\\2&0\end{pmatrix*},~~\text{and}~\sigma_z=\begin{pmatrix*}[r]1&0\\0&-1\end{pmatrix*},
\end{equation}
and $\hat{\vec{S}}\leftrightarrow\frac{1}{2}\vec{\sigma}$ is the faithful representation of spin-$\frac{1}{2}$. Using the Pauli matrix identity
\begin{equation}
    e^{i\vec{v}\cdot\vec{\sigma}}=\cos|\vec{v}|\mathbb{I}_2+i\sin|\vec{v}|\frac{\vec{v}}{|\vec{v}|}\cdot\vec{\sigma},
\end{equation}
with $\vec{v}$ a real-valued three-vector, we can compute
\begin{equation}
    e^{-i\theta\sigma_x}=\begin{pmatrix*}[r]\cos\theta&-i\sin\theta\\-i\sin\theta&\cos\theta\end{pmatrix*}.
    \label{eq:pauli1}
\end{equation}
Our goal is to rewrite this matrix exponential as the product of three matrix exponentials given by $e^{a\sigma_+}e^{b\sigma_z}e^{c\sigma_-}$. Substituting in the matrix exponentials of these Paulis gives
\begin{equation}
    e^{a\sigma_+}e^{b\sigma_z}e^{c\sigma_-}=\begin{pmatrix}1&2a\\0&1\end{pmatrix}\,\begin{pmatrix}e^b&0\\0&e^{-b}\end{pmatrix}\,\begin{pmatrix}1&0\\2c&1\end{pmatrix}=\begin{pmatrix}e^b+4ace^{-b}&2ae^{-b}\\2ce^{-b}&e^{-b}\end{pmatrix}.
    \label{eq:pauli2}
\end{equation}
Equating the right hand sides of Eqs.~(\ref{eq:pauli1}) and (\ref{eq:pauli2}), gives us
\begin{equation}
    a=-\frac{i}{2}\tan\theta,~~b=-\ln(\cos\theta),~~\text{and}~~c=-\frac{i}{2}\tan\theta.
\end{equation}
Rewriting in terms of the spin operators yields the exponential disentangling identity
\begin{align}
    e^{-i\theta(\hat{S}_++\hat{S}_-)}&=e^{-i\tan\theta\hat{S}_+}e^{-2\ln(\cos\theta)\hat{S}_z}e^{-i\tan\theta\hat{S}_-}\nonumber\\
    &=e^{-i\tan\theta\hat{S}_+}e^{-\ln(\cos\theta)(\hat{S}_+\hat{S}_--\hat{S}_-\hat{S}_+)}e^{-i\tan\theta\hat{S}_-}.
    \label{eq:disentangle}
\end{align}
This separates out the excitation operators to the left, the difference of the two projection operators in the center and the de-excitation operators to the right, for each UCC factor. Now, one might be concerned that the final matrices in the factorization are no longer unitary, so we have derived an identity starting from SU(2) but ending in a different group. This is indeed correct. The actual group we are working in for the disentangling identity is SL(2,$\mathbb{C}$), the special linear group of $2\times 2$ matrices with complex coefficients. This is also a Lie group and SU(2) is a subgroup of it, so this is the reason why the disentangling identity can be extended to  operators. But, if you are in doubt about this, we next verify it directly in terms of the operators.

Start from the fact that $e^{-i\tan\theta\hat{S}_\pm}=1-i\tan\theta\hat{S}_\pm$. Then expand
\begin{equation}
    e^{-2\ln(\cos\theta)\hat{S_z}}=\sum_{n=0}^\infty\frac{\Big (-2\ln(\cos\theta)\Big )^n}{n!}\Big (\hat{S}_z\Big )^n
\end{equation}
and note that 
\begin{equation}
\Big(2\hat{S}_z\Big)^2=(\hat{S}_+\hat{S}_--\hat{S}_-\hat{S}_+)^2=\hat{S}_+\hat{S}_-+\hat{S}_+\hat{S}_-, 
\end{equation}
so we have
\begin{align}
    e^{-2\ln(\cos\theta)\hat{S_z}}&=\hat{\mathbb{I}}+\Big(\cosh\Big(\ln(\cos\theta)\Big)-1\Big)(\hat{S}_+\hat{S}_-+\hat{S}_-\hat{S}_+)\nonumber\\
    &-\sinh\Big(\ln(\cos\theta)\Big)(\hat{S}_+\hat{S}_--\hat{S}_-\hat{S}_+).
\end{align}
Now, we substitute into the disentangling identity to find
\begin{align}
    e^{-i\theta(\hat{S}_++\hat{S}_-)}&=e^{-2\ln(\cos\theta)\hat{S_z}}-i\tan\theta\hat{S}_+e^{-2\ln(\cos\theta)\hat{S_z}}-ie^{-2\ln(\cos\theta)\hat{S_z}}\tan\theta\hat{S}_-\nonumber\\
    &-\tan^2\theta\hat{S}_+e^{-2\ln(\cos\theta)\hat{S_z}}\hat{S}_-\nonumber\\
    &=e^{-2\ln(\cos\theta)\hat{S_z}}-i\sin\theta\hat{S}_+-i\sin\theta\hat{S}_-+\tan\theta\sin\theta\hat{S}_+\hat{S}_-\nonumber\\
    &=\hat{\mathbb{I}}-i\sin\theta\hat{S}_+-i\sin\theta\hat{S}_-+(\cos\theta-1)\hat{S}_-\hat{S}_++\left(\frac{1-\sin^2\theta}{\cos\theta}-1\right)\hat{S}_+\hat{S}_-\nonumber\\
    &=\hat{\mathbb{I}}-i\sin\theta(\hat{S}_++\hat{S}_-)+(\cos\theta-1)(\hat{S}_+\hat{S}_-+\hat{S}_-\hat{S}_+),
\end{align}
which establishes the identity directly in terms of the operators. Note that because the operators in the factorization are not unitary, this factorization is not useful for creating quantum circuits on a quantum computer, it really is only useful for relating UCC to CC. In addition, note that the transformation is well-defined for $-\frac{\pi}{2}<\theta<\frac{\pi}{2}$, which is where we restrict all of the UCC amplitudes to lie.

The case $\theta=\pm\frac{\pi}{2}$ would present a problem for the disentangling identity due to divergences. But, in general, we do not expect there to ever be a UCC factor with $\theta=\frac{\pi}{2}$, because in such a case, the reference state is \textit{removed} from the wavefunction, when this term is applied ($\cos\theta=0$ implies no reference state remains, as shown by the Euler formula) . Since the reference is supposed to be a large amplitude term in the final wavefunction, we would never use such a large angle. In this work, we assume that none of the UCC amplitudes ever have a magnitude as large as $\frac{\pi}{2}$; usually, the magnitudes are not larger than $\frac{\pi}{4}$.

We want to make one additional comment about the disentangling identity. Note, how the amplitudes for the UCC are restricted to lie between $-\pi/2\le\theta\le\pi/2$, whereas CC amplitudes can have any real value. This is exhibited clearly in the disentangling identity, as an angle (which becomes a cosine or a sine using the $SU(2)$ identity), is replaced by a $\tan\theta$ in the disentangling identity, showing it can take any value. We wanted to stress this observation, because it makes sense as to why the identity has the form it has. Finally, because we anticipate most angles have their magnitude bound by $\pi/4$ instead of $\pi/2$, this indicates that in a CC approximation, most amplitudes should be less than 1 in magnitude.

The way the disentangling identity is used in the factorized form of the UCC is that we replace each UCC factor of the form $e^{-i\theta(\hat{S}_++\hat{S}_-)}$, with the second line of Eq.~(\ref{eq:disentangle}). This separates the excitation operators from the de-excitation operators, but it is not in the normal-ordered form (with all $\hat{S}_+$ to the left and all $\hat{S}_-$ to the right; we move the projection operator to the right, to the extent that we can, as well). In this normal ordered form, we have only the CC operators left. Note that all other operators will annihilate against $|\psi_0\rangle$. Recall as well that the spin operators depend on each specific excitation and de-excitation operator in the UCC factors, so one must convert from the spin notation back to the fermionic creation and destruction operators before putting the product of operators into normal-ordered form.

So, our next step is to determine how to re-order different exponential factors. This is done with the Hadamard lemma, which reads 
\begin{equation}
    e^{\hat{A}}\hat{B}e^{-\hat{A}}=\hat{B}+\frac{1}{1!}[\hat{A},\hat{B}]+\frac{1}{2!}[\hat{A},[\hat{A},\hat{B}]]+\frac{1}{3!}[\hat{A},[\hat{A},[\hat{A},\hat{B}]]]+\cdots,
\end{equation}
with the $n$th term being an $n$-fold nested commutator of $\hat{A}$ operators with one $\hat{B}$ operator on the right. The Hadamard lemma is often called the Baker-Campbell-Hausdorff formula in Chemistry literature. Because $e^{\hat{A}}\hat{B}^ne^{-\hat{A}}=\left (e^{\hat{A}}\hat{B}e^{-\hat{A}}\right )^n$, the Hadamard identity ``reaches inside'' the argument of functions. This means we have the exponential re-ordering identity
\begin{equation}
    e^{\hat{A}}e^{\hat{B}}=e^{e^{\hat{A}}\hat{B}e^{-\hat{A}}}e^{\hat{A}},
\end{equation}
after using the Hadamard lemma and multiplying the left and right sides by $e^{\hat{A}}$.

In principle, it is straightforward now to evaluate all of the re-orderings, but it is quite cumbersome to do so. The notation needed to describe these commutations is also cumbersome, so we need to develop new notation in order to carry out the re-orderings. Every term that we work with has the form of some number of raising operator factors and some number of lowering operator factors, always organized so the raising operators are to the left of the lowering operators, except in the projection operators, where half of the terms are in the opposite order. But unlike the original operators where the raising and lowering operators are always grouped into virtual or real spin-orbitals, after we have re-ordered exponentials, we will generate operators where the indices are mixed between the two groups. This is why we need a more general notation to describe the re-orderings.

We use the notation
\begin{align}
    \hat{A}(a_1,\cdots,a_n;b_1,\cdots,b_n|c_1,\cdots,c_m;d_1,\cdots,d_{m'})&=\underbrace{\hat{a}_{a_1}^\dagger\cdots\hat{a}_{a_n}^\dagger}_{n~\text{terms}}\,\underbrace{\hat{a}_{b_n}^{\phantom{\dagger}}\cdots\hat{a}_{b_1}^{\phantom{\dagger}}}_{n~\text{terms}}\,\underbrace{\hat{n}_{c_1}\cdots\hat{n}_{c_m}}_{m~\text{terms}}\nonumber\\
    &\times\underbrace{(1-\hat{n}_{d_1})\cdots(1-\hat{n}_{d_{m'}})}_{m'~\text{terms}},
\end{align}
to describe the states that we need to work with. The general operator is broken into two halves. The left half includes the isolated creation and annihilation operators, while the right half includes the projection operators---the number of creation operators is always the same as the number of destruction operators---hence the total number of creation plus destruction operators in each product string is even; the ordering of the destruction operators is reversed, just as is done in the excitation and de-excitation operators. Note that it is best to represent $\hat{n}$ as $\hat{a}^\dagger\hat{a}$ and $1-\hat{n}$ as $\hat{a}\hat{a}^\dagger$, so that each operator $\hat{A}$ is a product of creation and destruction operators, but it is often not normal ordered (for example, when $n\ne 0$ and $m$ or $m'$ are nonzero).

This operator can describe a \textit{pure excitation} when $m=m'=0$ and $\{a_1,\cdots,a_n\}$ are all virtual spin orbitals and $\{b_1,\cdots,b_n\}$ are all real spin orbitals. It is a \textit{pure de-excitation} when $m=m'=0$ and $\{a_1,\cdots,a_n\}$ are all real spin orbitals and $\{b_1,\cdots,b_n\}$ are all virtual spin orbitals. Similarly, when $n=0$ and $m\ne 0$ and/or $m'\ne 0$, the operator is a \textit{pure projection operator}. If the operator is not a pure operator, it is a \textit{mixed operator}. Such a mixed operator mixes excitation and de-excitation between different spin orbitals in the same operator. In cases when $m\ne 0$ and/or $m'\ne 0$, the operator is said to be \textit{with projection}. If $m=m'=0$ it is \textit{without projection}.

It is useful to describe some rules about these operators. If any $a_i=a_j$, any $b_i=b_j$, any $a_i=c_j$, any $b_i=d_j$, or any $c_i=d_j$, then the operator vanishes because it has the square of a creation or destruction operator or a product $\hat{n}(1-\hat{n})=0$. If any $c$ or $d$ index repeats, the repeating index can be removed. If any $a_i=d_j$, the $d_j$ index can be removed (because $\hat{a}_i^\dagger(1-\hat{n}_i)=\hat{a}_i^\dagger$) and if any $b_i=c_j$, the $c_j$ index can be removed (because $\hat{a}_i^{\phantom{\dagger}}\hat{n}_i=\hat{a}_i^{\phantom{\dagger}}$). Finally, if any $a_i=b_j$, we move it into the $c$ indices. We say the operator is in \textit{canonical form} if all of these ``contractions'' have been applied to the operator. In this case, all indices that are in the operator are different. In addition, when in canonical form, all of the indices in each grouping $a$, $b$, $c$, and $d$ has the index values in each subgroup ordered, so that $a_1<a_2<a_3\cdots$ and similarly for the other three sets of indices.

Before we start re-ordering the exponential factors to place the product in an ``excitation-only'' form, every operator in the exponential is a pure excitation without projection, a pure de-excitation without projection, or a pure projection, because this is the form of the exponential disentangling identity for each term. As we re-order operators, this changes, and many operators become mixed and with projection. But, there are some simple rules for how the re-ordering changes terms, which we go through next. These rules are rather tedious to carry out ``by hand,'' but are straightforward to implement on a computer.

The re-ordering of exponential terms always involves interchanging the order of an $\hat{A}$ operator and an $\hat{A}'$ operator, in the form
\begin{equation}
    e^{\alpha\hat{A}}e^{\alpha'\hat{A}'}=e^{e^{\alpha\hat{A}}\alpha'\hat{A}'e^{-\alpha\hat{A}}}e^{\alpha\hat{A}},
\end{equation}
where the operators $\hat{A}$ and $\hat{A}'$ can both be assumed to be in canonical form.
In most cases, the Hadamard lemma truncates after a finite number of nested commutators, for the following reasons. First, the commutator vanishes if  $\{a_i\mid i=1,\cdots,n\}\cap\{a_i^\prime\mid i=1,\cdots,n'\}\ne0$ or $\{b_i\mid i=1,\cdots,n\}\cap\{b_i^\prime\mid i=1,\cdots,n'\}\ne0$ and no $a_i$ index is in $\{b_i^\prime\}$ and no $b_i$ is in $\{a_i^\prime\}$, because if  there is a common index in both operators, the canonical form guarantees there is no $c$ or $d$ index that is the same, so both $\hat{A}$ and $\hat{A}'$ have the same creation operator in them, or the same destruction operator in them, but not both. This means their product vanishes in either order (because the square of a fermionic operator is zero). So, we assume these two sets have no intersection.
Then the commutator $[\hat{A},\hat{A}']\ne 0$ if and only if there is at least one of the following: 
\begin{align}
    &\exists ~a_i=b_j^\prime~~\text{or}~~c_j^\prime~~\text{or}~~d_j^\prime\label{eq:case1}\\
    &\exists ~b_i=a_j^\prime~~\text{or}~~c_j^\prime~~\text{or}~~d_j^\prime.\label{eq:case2}\\
    &\exists ~c_i=a_j^\prime~~\text{or}~~b_j^\prime\label{eq:case3}\\
    &\exists ~d_i=a_j^\prime~~\text{or}~~b_j^\prime\label{eq:case4}
\end{align}
Because the operators are in canonical form, only one of the possibilities (for any specific $i$) in any line can occur, but there can be more than one $i$ that satisfies this condition for any line and we can have the condition satisfied on more than one line.

Let us carefully look at the cases. If Eq.~(\ref{eq:case1}) holds, then if the match is with $c_j^\prime$ or $d_j^\prime$, every term in the commutator has an $a_{a_i}^\dagger$ factor in the product of operators (after putting the commutator into canonical form), which annihilates with $\hat{A}$ when multiplied on either side. This means only the first commutator is nonzero and all higher-order ones vanish. The same is true for Eq.~(\ref{eq:case2}) if the match is with  $c_j^\prime$ or $d_j^\prime$. 
If Eq.~(\ref{eq:case3}) holds we have two different possible behaviors. First, if $\{a_i\mid i=1,\cdots,n\}\subset\{a_i^\prime\mid i=1,\cdots,n'\}$ and $\{b_i\mid i=1,\cdots,n\}\subset \{b_i^\prime\mid i=1,\cdots,n'\}$, then in each term in the commutator, all of the unpaired creation and annihilation operators in $\hat{A}$ are paired into $\hat{n}$ or $1-\hat{n}$ factors. In this case, the nested commutators go on for an infinite number of terms, but they can all be summed into an exponential factor that multiplies the result. But, if both sets for the indices of the creation and destruction operators of $\hat{A}$ are not also in the labels for $\hat{A}'$, then we always have at least one creation or annihilation operator left over in every term in the commutator---in this situation, $\hat{A}$ annihilates the operator when multiplied to the left or the right, and the expansion truncates after the first commutator. We have a similar result for Eq.~(\ref{eq:case4}).
The remaining case occurs when the match is only between the creation and annihilation operators in Eqs.~(\ref{eq:case1}) and (\ref{eq:case2}).

In this case, the number operators commute with everything, so we need not worry about them anymore. Then we have at most two nested commutators. Consider the first index $i$, such that $a_i=b_j^\prime$. After the first commutator, one term in the commutator has no $a_{a_i}^\dagger$ and $\hat{a}_{a_i}^{\phantom{\dagger}}$ terms in it, while the rest have either a $\hat{n}_{a_i}$ term or a $1-\hat{n}_{a_i}$ term. After the second nested commutator, all terms will have a $\hat{a}_{a_i}$ term in it, which then vanishes when multiplied by $\hat{A}$ on either side for the third nested commutator. We have a similar argument if the only indices that are the same are $b_i=a_j^\prime$. The second commutator only enters in situations where both $\{a_i\mid i=1,\cdots,n\}\subset\{b_i^\prime\mid i=1,\cdots,n'\}$ and $\{b_i\mid i=1,\cdots,n\}\subset \{a_i^\prime\mid i=1,\cdots,n'\}$, as before, since, otherwise, we have a lone fermionic operator remaining in all terms of the commutator, which then vanishes when multiplied by $\hat{A}$ on the right or on the left.
So, the reordering is always of the form
\begin{equation}
    e^{\alpha\hat{A}}e^{\alpha'\hat{A}'}=e^{\alpha'\hat{A}'+\alpha\alpha'[\hat{A},\hat{A}']}e^{\alpha\hat{A}},
\end{equation}
when the $a$ and $b$ indices of the operator $\hat{A}$ do not entirely lie within the $b'$ and $a'$ indices of $\hat{A}'$. We use the following terminology: when the $a$ and $b$ indices of $\hat{A}$ are a subset of the $b'$ and $a'$ indices of $\hat{A}'$, we say $\hat{A}$ \textit{matches} $\hat{A}'$; if they are not both subsets, we say $\hat{A}$ \textit{does not match} $\hat{A}'$.

If $\hat{A}'$ does not match $\hat{A}$, then the commutator $[\hat{A},\hat{A}']$ has lone creation or annihilation operators from $\hat{A}'$ in it, meaning $\hat{A}'$ annihilates it when multiplied to the right or to the left (after putting the commutator into canonical form). Hence, $\hat{A}'$ and $[\hat{A},\hat{A}']$ commute and we can separate the factors into
\begin{equation}
    e^{\alpha\hat{A}}e^{\alpha'\hat{A}'}=e^{\alpha'\hat{A}'}e^{\alpha\alpha'[\hat{A},\hat{A}']}e^{\alpha\hat{A}}.
    \label{eq:reorder1}
\end{equation}
The re-ordering has now been put into a form of exponentials of individual A-form operators. If $\hat{A}'$ matches $\hat{A}$, then the commutator $[\hat{A}',[\hat{A},\hat{A}']]$ has lone creation or annihilation operators in it (following similar arguments as given above), so it commutes with $\hat{A}'$, but it does not necessarily commute with $[\hat{A},\hat{A}']$ (which has no lone fermionic operators from $\hat{A}'$). In this case, there is no simple way to separate the terms, as nested commutators of arbitrary order will be nonzero. But, of course, it can be put into an infinite product of exponentials of individual A-form operators, by use of the Zassenhaus formula (and many of the nested commutators in that formula will vanish). Because the commutator of a projection operator with a fermionic operator is proportional to the fermionic operator, the nested commutators will repeat in form, and in the end, these infinite factors can all be combined with a new numerical factor multiplying the corresponding operator. To illustrate how this might work, we use the exact formula for the Baker-Campbell-Hausdorff formula~\cite{jacobsen,pascual}
\begin{align}
    e^{\hat{A}}e^{\hat{B}}&=e^{\eta(\hat{A},\hat{B})},\\
    \eta(\hat{A},\hat{B})&=\sum_{m=1}^\infty\frac{(-1)^{m-1}}{m}\nonumber\\
    &\times\sum_{p_i\ge 0,~q_i\ge 0,~p_i+q_i\ge 1}\frac{[\overbrace{\hat{A}\cdots\hat{A}}^{p_1}\overbrace{\hat{B}\cdots\hat{B}}^{q_1}\cdots\overbrace{\hat{A}\cdots\hat{A}}^{p_m}\overbrace{\hat{B}\cdots\hat{B}}^{q_m}]}{\left (\sum_j(p_j+q_j)\right )p_1!q_1!\cdots p_m!q_m!}\\
    [\hat{C}\hat{D}\hat{E}\cdots\hat{J}]&=[\cdots[[\hat{C},\hat{D}],\hat{E},\cdots,\hat{J}].
\end{align}
For example, the first few terms correspond to $m=1$, $\sum_i(p_i+q_i)=1$ ($p_1=1$, $q_1=0$ and $p_1=0$, $q_1=1$), $m=1$, $\sum_i(p_i+q_i)=2$ ($p_1=2$, $q_1=0$, and $p_1=1$, $q_1=1$, and $p_1=0$, $q_1=2$), and $m=2$, $\sum_i(p_i+q_i)=2$ ($p_1=1$, $q_1=0$, $p_2=1$, $q_2=0$, and $p_1=1$, $q_1=0$, $p_2=0$, $q_2=1$, and $p_1=0$, $q_1=1$, $p_2=1$, $q_2=0$, and $p_1=0$, $q_1=1$, $p_2=0$, $q_2=1$), and so on. We must take into account all possibilities for the formula, but many of the commutators vanish (for example, $p_1=2$, $q_1=0$ vanishes and all $m=2$, $\sum(p+q)=2$ terms vanish as well). The net contribution becomes
\begin{equation}
\eta(\hat{A},\hat{B})=\underbrace{\hat{A}+\hat{B}}_{m=1,~\sum(p+q)=1}+\underbrace{\frac{1}{2}[\hat{A},\hat{B}]}_{m=1,~\sum(p+q)=2}+\underbrace{\frac{1}{12}[[\hat{A},\hat{B}],\hat{B}]+\frac{1}{12}[[\hat{B},\hat{A}],\hat{A}]}_{m=2,~\sum(p+q)=3}+\cdots.
\end{equation}
Now, back to the problem at hand, involving separating the two terms in Eq.~(\ref{eq:reorder1}) when $\hat{A}'$ matches $\hat{A}$. If we let $\hat{A}'\to \hat{A}$ and $[\hat{A},\hat{A}']\to \hat{B}$ in the Baker-Campbell-Hausdorff formula, then, because $[\hat{A},[\hat{A},\hat{B}]]=0$, all terms with $p_i\ge 2$ vanish in the expansion. This will simplify the expression, but it is not enough for us to determine the required Zassenhaus factorization of the expression, without significant additional computation.

Now, if $\hat{A}$ does match $\hat{A}'$, then the exponential re-ordering identity takes the form
\begin{equation}
    e^{\alpha\hat{A}}e^{\alpha'\hat{A}'}=e^{\alpha'\hat{A}'+\alpha\alpha'[\hat{A},\hat{A}']+\frac{1}{2}\alpha^2\alpha'[\hat{A},[\hat{A},\hat{A}']]}e^{\alpha\hat{A}}.
\end{equation}
If $\hat{A}'$ does not match $\hat{A}$, then we have lone fermionic operators from $\hat{A}'$ in each of the commutators, which means the three separate operator terms in the exponent, all mutually commute, and we can re-write the exponential re-ordering identity as
\begin{equation}
    e^{\alpha\hat{A}}e^{\alpha'\hat{A}'}=e^{\alpha'\hat{A}'}e^{\alpha\alpha'[\hat{A},\hat{A}']}e^{\frac{1}{2}\alpha^2\alpha'[\hat{A},[\hat{A},\hat{A}']]}e^{\alpha\hat{A}}.
\end{equation}
The case where $\hat{A}'$ also matches $\hat{A}$ is again complicated and we have to use the infinite-order Zassenhaus formula again, perhaps recursively, because there are three terms in the exponent. In this case, the operators may not reduce to a simple form.

Finally, we still have to work out the cases from Eqs.~(\ref{eq:case3}) and $(\ref{eq:case4})$ when $\hat{A}$ matches $\hat{A}'$. To do this, we first need to determine the general similarity transformation
\begin{align}
    \hat{\mathcal{O}}(\alpha)=\exp\left ({\alpha\prod_{i=1}^{m+n}\hat{n}_{i}\prod_{i=m+n+1}^{m+n+m'+n'}(1-\hat{n}_i)}\right )&\left (
    \prod_{i=1}^m\hat{a}_i^\dagger\prod_{i=m+1}^{m+n}\hat{a}_i^{\phantom{\dagger}}\prod_{i=m+n+1}^{m+n+m'}\hat{a}_i^\dagger\prod_{i=m+n+m'+1}^{m+n+m'+n'}\hat{a}_i^{\phantom{\dagger}}\right )\nonumber\\
    &\times \exp\left ({-\alpha\prod_{i=1}^{m+n}\hat{n}_{i}\prod_{i=m+n+1}^{m+n+m'+n'}(1-\hat{n}_i)}\right ),
\end{align}
which is most easily calculated by computing the derivative with respect to $\alpha$. This creates a commutator inside the similarity transformation, which is best evaluated by directly multiplying the projection operators from the left and from the right. Immediately, we see that the commutator vanishes unless $m=n'=0$ or $n=m'=0$.  Hence, we find
\begin{equation}
\frac{d}{d\alpha}\hat{\mathcal{O}}(\alpha)=\begin{cases}
\phantom{-}\hat{\mathcal{O}}(\alpha)~&\text{if}~n=m'=0\\
-\hat{\mathcal{O}}(\alpha)~&\text{if}~m=n'=0\\
~~~~~~0~&\text{otherwise.}\end{cases}
\label{eq:dif-O}
\end{equation}
This means $\hat{\mathcal{O}}(\alpha)$ is either unchanged, or multiplied by a factor of $e^{\pm\alpha}$ after the re-ordering of the operators, depending on the cases listed in Eq.~(\ref{eq:dif-O}). As we will see in the examples below, this correction typically changes an amplitude by a factor of a secant of another amplitude; in many cases the secant is raised to an operator exponent.

We need to have a brief summary about this complex algebra. What we have found is that in cases where neither $\hat{A}$ matches $\hat{A}'$ nor $\hat{A}'$ matches $\hat{A}$, the exponential re-ordering is simple because it involves evaluating one commutator and requires no additional operator manipulations. If $\hat{A}$ matches $\hat{A}'$, but $\hat{A}'$ does not match $\hat{A}$, then we have two  commutators to evaluate. But if both operators match in both directions, the formula is complicated, and might not even be able to be written down analytically. This raises the question, can such a situation occur? In general, it seems like it is a rare occurrence, if it happens at all, but without looking at some examples we will not be able to tell. In cases where this does not occur, we have established the rules needed to perform these algebraic manipulations, but they are clearly complicated to try to carry out ``by hand.'' However, they are straightforward to implement using computer algebra---but, this will not be done in this work. We argue why regardless, the final formulas should be able to be worked out later.

\section{Examples of the connection between the factorized form of UCC and CC}

We begin by examining the special case where all excitations and de-excitations are singles operators, of the form $\theta_{i}^{a}(\hat{a}_{a}^{\dagger}\hat{a}_{i}^{\phantom{\dagger}}-\hat{a}_{i}^{\dagger}\hat{a}_{a}^{\phantom{\dagger}})$. The reason why this is a simple case is that singles excitation and de-excitation operators form a closed Lie algebra amongst themselves~\cite{izmaylov}, so, in principle, the exponential re-ordering does not explode into many high-rank operators. Let us see how this happens.

The exponential disentangling identity for the singles operator is
\begin{equation}
    e^{\theta(\hat{a}_{a}^{\dagger}\hat{a}_{i}^{\phantom{\dagger}}-\hat{a}_{i}^{\dagger}\hat{a}_{a}^{\phantom{\dagger}})}=e^{\tan\theta\,\hat{a}_{a}^{\dagger}\hat{a}_{i}^{\phantom{\dagger}}}e^{-\ln\big(\cos\theta\big)[\hat{n}_a(1-\hat{n}_i)-\hat{n}_i(1-\hat{n}_a)]}e^{-\tan\theta\,\hat{a}_{i}^{\dagger}\hat{a}_{a}^{\phantom{\dagger}}}.
\end{equation}
While the identities we discuss hold at the operator level, since we are interested in the connection between CC and UCC, it is more convenient to act the operator onto the initial reference state, because this allows us to remove many factors from the final operator. We start with two factors and show how we disentangle them and relate them back to the CC operator. So we have, using our general notation,
\begin{align}
    e^{\theta_j^b(\hat{a}_b^\dagger\hat{a}_j^{\phantom{\dagger}}-\hat{a}_j^\dagger\hat{a}_b^{\phantom{\dagger}})}e^{\theta_i^a(\hat{a}_a^\dagger\hat{a}_i^{\phantom{\dagger}}-\hat{a}_i^\dagger\hat{a}_a^{\phantom{\dagger}})}|\Psi_0\rangle&=e^{\tan\theta_j^b\hat{A}(b;j|)}e^{-\ln\big (\cos\theta_j^b\big)\big(\hat{A}(|b;j)-\hat{A}(|j;b)\big)}e^{-\tan\theta_j^b\hat{A}(j;b|)}\nonumber\\
    &\times e^{\tan\theta_i^a\hat{A}(a;i|)}e^{-\ln\big (\cos\theta_i^a\big)\big(\hat{A}(|a;i)-\hat{A}(|i;a)\big)}\cancel{e^{-\tan\theta_i^a\hat{A}(i;a|)}}|\Psi_0\rangle,
\end{align}
where we leave empty the indices in $\hat{A}$ that have no corresponding operators in $\hat{A}$.
We also have cancelled the last term, since the exponent annihilates the reference state. The middle term on the lower line evaluates to $\cos\theta_i^a$ against the reference. We need to re-order the leftmost operator on the last line through the two operators to its left. We start with re-ordering
\begin{equation}
    e^{-\tan\theta_j^b\hat{A}(j;b|)}e^{\tan\theta_i^a\hat{A}(a;i|)}=e^{\tan\theta_i^a\hat{A}(a;i|)}
    e^{\tan\theta_i^a\tan\theta_j^b(\big(\delta_{ij}\hat{A}(a;b|)-\delta_{ab}\hat{A}(j;i|)\big)}
    e^{-\tan\theta_j^b\hat{A}(j;b|)}.
\end{equation}
The last operator is removed when it acts on the reference. The middle operator (which is a mixed operator) can only have one term in it, since we cannot have both $i=j$ and $a=b$, otherwise the two original UCC factors would have been identical, which we assume does not happen. In either case, when this middle term operates on the reference state, the exponent annihilates, because we cannot destroy in the $b$ spin-orbital nor create in the $j$ spin-orbital. So, we are left with the second re-ordering we need to do, namely
\begin{align}
    e^{-\ln\big (\cos\theta_j^b\big)\big(\hat{A}(|b;j)-\hat{A}(|j;b)\big)}e^{\tan\theta_i^a\hat{A}(a;i|)}&=e^{\tan\theta_i^a\hat{A}(a;i|)}
    e^{-\tan\theta_i^a\ln\big(\cos\theta_j^b\big)(\delta_{ij}+\delta_{ab})\hat{A}(a;i|)}\nonumber\\
    &\times
    e^{-\ln\big (\cos\theta_j^b\big)\big(\hat{A}(|b;j)-\hat{A}(|j;b)\big)}.
\end{align}
The rightmost term gives a factor of $\cos\theta_j^b$, while the middle term combines with the left most term, since the operators are the same.
So, after the re-ordering, we have
\begin{align}
    e^{\theta_j^b(\hat{a}_b^\dagger\hat{a}_j^{\phantom{\dagger}}-\hat{a}_j^\dagger\hat{a}_b^{\phantom{\dagger}})}e^{\theta_i^a(\hat{a}_a^\dagger\hat{a}_i^{\phantom{\dagger}}-\hat{a}_i^\dagger\hat{a}_a^{\phantom{\dagger}})}|\Psi_0\rangle&=\cos\theta_j^b\cos\theta_i^a e^{\tan\theta_j^b\hat{a}_b^\dagger\hat{a}_j^{\phantom{\dagger}}+\tan\theta_i^a\big(1-\ln\big(\cos\theta_j^b\big)(\delta_{ij}+\delta_{ab})\big)\hat{a}_a^\dagger\hat{a}_i^{\phantom{\dagger}}}|\Phi_0\rangle.
\end{align}
This is in the CC form, with one exponential of a sum of excitation operators. Note that if none of the indices are the same, it gives us the standard form. You can also check that if either $i=j$ or $a=b$, where the double-excitation term cannot be excited, we also get the correct state. We can also see that the relationship between the two ansatzes is complicated. A UCC amplitude has to have its tangent taken for the CC amplitude. The overall factor of cosines is neglected in the traditional CC ansatz, which is why it does not preserve the norm of the state, in most cases.

This example does not show the effect of de-excitation, because neither of the terms given there can de-excite, yet. But if we add in an additional singles factor, removing the spin-orbital $k$ and occupying the spin-orbital $c$, then we will have a de-excitation if $c=a$ and $k=j$, for example. So we consider this case next. We have $i\ne j$ and $a\ne b$. Then
\begin{align}
    e^{\theta_j^a(\hat{a}_a^\dagger\hat{a}_j^{\phantom{\dagger}}-\hat{a}_j^\dagger\hat{a}_a^{\phantom{\dagger}})}&
    e^{\theta_j^b(\hat{a}_b^\dagger\hat{a}_j^{\phantom{\dagger}}-\hat{a}_j^\dagger\hat{a}_b^{\phantom{\dagger}})}e^{\theta_i^a(\hat{a}_a^\dagger\hat{a}_i^{\phantom{\dagger}}-\hat{a}_i^\dagger\hat{a}_a^{\phantom{\dagger}})}|\Psi_0\rangle=\cos\theta_j^b\cos\theta_i^a
    e^{\tan\theta_j^a\hat{a}_{a}^{\dagger}\hat{a}_{j}^{\phantom{\dagger}}}
    \nonumber\\
    &\times
    e^{-\ln\big(\cos\theta_j^a\big)[\hat{n}_a(1-\hat{n}_j)-\hat{n}_j(1-\hat{n}_a)]} e^{-\tan\theta_j^a\hat{a}_{j}^{\dagger}\hat{a}_{a}^{\phantom{\dagger}}} e^{\tan\theta_j^b\hat{a}_b^\dagger\hat{a}_j^{\phantom{\dagger}}+\tan\theta_i^a\hat{a}_a^\dagger\hat{a}_i^{\phantom{\dagger}}}|\Phi_0\rangle.
\end{align}
We have two more re-orderings to carry out. The first is
\begin{align}
    &e^{-\tan\theta_j^a\hat{a}_{j}^{\dagger}\hat{a}_{a}^{\phantom{\dagger}}} e^{\tan\theta_j^b\hat{a}_b^\dagger\hat{a}_j^{\phantom{\dagger}}}e^{\tan\theta_i^a\hat{a}_a^\dagger\hat{a}_i^{\phantom{\dagger}}}\nonumber\\
    &~~~~~=e^{\tan\theta_j^b\hat{a}_b^\dagger\hat{a}_j^{\phantom{\dagger}}+\tan\theta_j^a\tan\theta_j^b\hat{a}_b^\dagger\hat{a}_a^{\phantom{\dagger}}}e^{-\tan\theta_j^a\hat{a}_{j}^{\dagger}\hat{a}_{a}^{\phantom{\dagger}}}e^{\tan\theta_i^a\hat{a}_a^\dagger\hat{a}_i^{\phantom{\dagger}}}\nonumber\\
    &~~~~~=e^{\tan\theta_j^b\hat{a}_b^\dagger\hat{a}_j^{\phantom{\dagger}}}e^{\tan\theta_j^a\tan\theta_j^b\hat{a}_b^\dagger\hat{a}_a^{\phantom{\dagger}}-\tan\theta_j^a\hat{a}_{j}^{\dagger}\hat{a}_{a}^{\phantom{\dagger}}}e^{\tan\theta_i^a\hat{a}_a^\dagger\hat{a}_i^{\phantom{\dagger}}}\nonumber\\
    &~~~~~=e^{\tan\theta_j^b\hat{a}_b^\dagger\hat{a}_j^{\phantom{\dagger}}}e^{\tan\theta_i^a\hat{a}_a^\dagger\hat{a}_i^{\phantom{\dagger}}+\tan\theta_j^a\tan\theta_j^b\tan\theta_i^a\hat{a}_b^\dagger\hat{a}_i^{\phantom{\dagger}}-\tan\theta_j^a\tan\theta_i^a\hat{a}_j^\dagger\hat{a}_i^{\phantom{\dagger}}}
    e^{\tan\theta_j^a\tan\theta_j^b\hat{a}_b^\dagger\hat{a}_a^{\phantom{\dagger}}-\tan\theta_j^a\hat{a}_{j}^{\dagger}\hat{a}_{a}^{\phantom{\dagger}}}\nonumber\\
    &~~~~~=e^{\tan\theta_j^b\hat{a}_b^\dagger\hat{a}_j^{\phantom{\dagger}}}e^{\tan\theta_i^a\hat{a}_a^\dagger\hat{a}_i^{\phantom{\dagger}}}e^{\tan\theta_j^a\tan\theta_j^b\tan\theta_i^a\hat{a}_b^\dagger\hat{a}_i^{\phantom{\dagger}}}e^{-\tan\theta_j^a\tan\theta_i^a\hat{a}_j^\dagger\hat{a}_i^{\phantom{\dagger}}}
    e^{\tan\theta_j^a\tan\theta_j^b\hat{a}_b^\dagger\hat{a}_a^{\phantom{\dagger}}}e^{-\tan\theta_j^a\hat{a}_{j}^{\dagger}\hat{a}_{a}^{\phantom{\dagger}}}.
\end{align}
The three rightmost terms are removed when they act on $|\Psi_0\rangle$.

The second is
\begin{align}
    &e^{-\ln\big(\cos\theta_j^a\big)[\hat{n}_a(1-\hat{n}_j)-\hat{n}_j(1-\hat{n}_a)]}e^{\tan\theta_j^b\hat{a}_b^\dagger\hat{a}_j^{\phantom{\dagger}}}e^{\tan\theta_i^a\hat{a}_a^\dagger\hat{a}_i^{\phantom{\dagger}}}e^{\tan\theta_j^a\tan\theta_j^b\tan\theta_i^a\hat{a}_b^\dagger\hat{a}_i^{\phantom{\dagger}}}\nonumber\\
    &~~~~~=e^{\tan\theta_j^b\sec\theta_j^a\hat{a}_b^\dagger\hat{a}_j^{\phantom{\dagger}}}
    e^{-\ln\big(\cos\theta_j^a\big)[\hat{n}_a(1-\hat{n}_j)-\hat{n}_j(1-\hat{n}_a)]}
    e^{\tan\theta_i^a\hat{a}_a^\dagger\hat{a}_i^{\phantom{\dagger}}}
    e^{\tan\theta_j^a\tan\theta_j^b\tan\theta_i^a\hat{a}_b^\dagger\hat{a}_i^{\phantom{\dagger}}}\nonumber\\
    &~~~~~=e^{\tan\theta_j^b\sec\theta_j^a\hat{a}_b^\dagger\hat{a}_j^{\phantom{\dagger}}}
    e^{\tan\theta_i^a\sec\theta_j^a\hat{a}_a^\dagger\hat{a}_i^{\phantom{\dagger}}}
    e^{\tan\theta_j^a\tan\theta_j^b\tan\theta_i^a\hat{a}_b^\dagger\hat{a}_i^{\phantom{\dagger}}}e^{-\ln\big(\cos\theta_j^a\big)[\hat{n}_a(1-\hat{n}_j)-\hat{n}_j(1-\hat{n}_a)]};
\end{align}
when the rightmost operator acts against the reference state, it produces a $\cos\theta_j^a$. Note that the exponential re-orderings here are of the form where the Hadamard has an infinite number of terms, but they can all be summed and the net effect is to renormalize the coefficients of some term by multiplying them by a power of the secant.

Putting them all together then gives us
\begin{align}
    e^{\theta_j^a(\hat{a}_a^\dagger\hat{a}_j^{\phantom{\dagger}}-\hat{a}_j^\dagger\hat{a}_a^{\phantom{\dagger}})}&
    e^{\theta_j^b(\hat{a}_b^\dagger\hat{a}_j^{\phantom{\dagger}}-\hat{a}_j^\dagger\hat{a}_b^{\phantom{\dagger}})}e^{\theta_i^a(\hat{a}_a^\dagger\hat{a}_i^{\phantom{\dagger}}-\hat{a}_i^\dagger\hat{a}_a^{\phantom{\dagger}})}|\Psi_0\rangle=\cos\theta_j^b\cos\theta_j^a\cos\theta_i^a
    \nonumber\\
    &\times e^{\tan\theta_j^a\hat{a}_{a}^{\dagger}\hat{a}_{j}^{\phantom{\dagger}}+\tan\theta_j^b\sec\theta_j^a\hat{a}_b^\dagger\hat{a}_j^{\phantom{\dagger}}+\tan\theta_i^a\sec\theta_j^a\hat{a}_a^\dagger\hat{a}_i^{\phantom{\dagger}}+\tan\theta_j^a\tan\theta_j^b\tan\theta_i^a\hat{a}_b^\dagger\hat{a}_i^{\phantom{\dagger}}}|\Psi_0\rangle.
\end{align}
Again, one can see there is a complicated relationship between the UCC amplitudes and the CC amplitudes. If we expand the operators using the two different identities (the generalization of the Euler formula versus the above result), we find the two wavefunctions agree, as they must. Note that we have to expand the exponential in the CC form through the quadratic power to include all terms.
This example is instructive, because it clearly shows that different amplitudes can enter into the CC than enter into the exponential factors in the factorized form of the UCC. This essentially underlies the earlier work that showed that a UCC and CC ansatz are not equivalent; of course, we should never have expected them to be. Our goal is to determine how they inter-relate. Indeed, it is most likely that a low-rank representation (say UCC-SD in a factorized form) will not be represented by a low-rank CC ansatz, and this is the key to the difference in the ansatzes.

In exploring these singles excitations, everything worked nicely, because the terms in the exponents always were of a similar form. In addition, for the cases we looked at, we never encountered the situation where $\hat{A}$ matched $\hat{A}'$ and $\hat{A}'$ matched $\hat{A}$, which is the case that is challenging for our operator identities.

Now, we will explore how the conversion from UCC to CC works for rank-two (doubles) terms. We start by re-ordering the case with two doubles UCC factors. First, we use the exponential disentangling identity to separate the terms in each:
\begin{align}
    &e^{\theta_{kl}^{cd}\big (\hat{a}_c^\dagger\hat{a}_d^\dagger\hat{a}_l^{\phantom{\dagger}}\hat{a}_k^{\phantom{\dagger}}-\hat{a}_k^\dagger\hat{a}_l^\dagger\hat{a}_d^{\phantom{\dagger}}\hat{a}_c^{\phantom{\dagger}}\big)}
    e^{\theta_{ij}^{ab}\big (\hat{a}_a^\dagger\hat{a}_b^\dagger\hat{a}_j^{\phantom{\dagger}}\hat{a}_i^{\phantom{\dagger}}-\hat{a}_i^\dagger\hat{a}_j^\dagger\hat{a}_b^{\phantom{\dagger}}\hat{a}_a^{\phantom{\dagger}}\big)}|\Psi_0\rangle\nonumber\\
    &~~~~~=e^{\theta_{kl}^{cd}\big(\hat{A}(cd;kl|)-\hat{A}(kl;cd|)\big)}e^{\theta_{ij}^{ab}\big(\hat{A}(ab;ij|)-\hat{A}(ij;ab|)\big)}|\Psi_0\rangle\nonumber\\
    &~~~~~=e^{\tan\theta_{kl}^{cd}\hat{A}(cd;kl|)}e^{-\ln\big(\cos\theta_{kl}^{cd}\big)\big(\hat{A}(|cd;kl)-\hat{A}(|kl;cd)\big)}e^{-\tan\theta_{kl}^{cd}\big(\hat{A}(kl;cd)\big)}\nonumber\\
    &~~~~~\times e^{\tan\theta_{ij}^{ab}\hat{A}(ab;ij|)}e^{-\ln\big(\cos\theta_{ij}^{ab}\big)\big(\hat{A}(|ab;ij)-\hat{A}(|ij;ab)\big)}e^{-\tan\theta_{ij}^{ab}\big(\hat{A}(ij;ab)\big)}|\Psi_o\rangle.
\end{align}
The rightmost term has an exponent that annihilates against the reference and can be removed. The second rightmost term will yield a cosine when it acts on the reference. Our first re-ordering involves the next two terms, as we move leftward through the products. It becomes
\begin{equation}
    e^{-\tan\theta_{kl}^{cd}\hat{A}(kl;cd)}
    e^{\tan\theta_{ij}^{ab}\hat{A}(ab;ij|)}=
    e^{\tan\theta_{ij}^{ab}\hat{A}(ab;ij|)-\tan\theta_{ij}^{ab}\tan\theta_{kl}^{cd}[\hat{A}(kl;cd),\hat{A}(ab;ij)]
    }
    e^{-\tan\theta_{kl}^{cd}\hat{A}(kl;cd)},
\end{equation}
because we cannot have $\hat{A}$ match $\hat{A}'$, because the two doubles operators would be identical if they did. These calculations get quite lengthy, as there are eight possible terms that can contribute in this first re-ordering. But because $a\ne b$, $c\ne d$, $i\ne j$, and $k\ne l$, at most three terms can contribute. However, it is cumbersome to include the most general situation, even for doubles. So, instead, we will consider some specific cases.  The first one we will look at is where one index is in common, and we choose, as an example the case $i=k$. This gives
\begin{equation}
    e^{-\tan\theta_{kl}^{cd}\hat{A}(kl;cd)}
    e^{\tan\theta_{ij}^{ab}\hat{A}(ab;ij|)}=
    e^{\tan\theta_{ij}^{ab}\hat{A}(ab;ij|)}e^{-\tan\theta_{ij}^{ab}\tan\theta_{kl}^{cd}\hat{A}(l,a,b;j,c,d|)
    }
    e^{-\tan\theta_{kl}^{cd}\hat{A}(kl;cd)},
\end{equation}
where we use a convention where the real orbital indices are less than the virtual orbital indices. Note how the commutator form is a triples operator. This is what typically happens with higher-rank operators---commutators increase the rank. The rightmost two terms annihilate against the reference state (this is true in the general case for all possible terms that can arise from the commutator). The second re-ordering is
\begin{align}
    e^{-\ln\big(\cos\theta_{il}^{cd}\big)\big(\hat{A}(|cd;il)-\hat{A}(|il;cd)\big)}e^{\tan\theta_{ij}^{ab}\hat{A}(ab;ij|)}&=e^{\tan\theta_{ij}^{ab}\big(\cos\theta_{il}^{cd}\big)^{-\hat{n}_l-\hat{n}_c+\hat{n}_l(\hat{n}_c+\hat{n}_d)}\hat{A}(ab;ij|)}\nonumber\\
    &\times e^{-\ln\big(\cos\theta_{il}^{cd}\big)\big(\hat{A}(|cd;il)-\hat{A}(|il;cd)\big)},
\end{align}
which requires the full Hadamard, just like before. When the right-most term operates on the reference, it gives a cosine as well. For the next term, we can replace the exponent by $-1$ when acting on the reference. Putting this all together yields
\begin{align}
    &e^{\theta_{kl}^{cd}\big (\hat{a}_c^\dagger\hat{a}_d^\dagger\hat{a}_l^{\phantom{\dagger}}\hat{a}_k^{\phantom{\dagger}}-\hat{a}_k^\dagger\hat{a}_l^\dagger\hat{a}_d^{\phantom{\dagger}}\hat{a}_c^{\phantom{\dagger}}\big)}
    e^{\theta_{ij}^{ab}\big (\hat{a}_a^\dagger\hat{a}_b^\dagger\hat{a}_j^{\phantom{\dagger}}\hat{a}_i^{\phantom{\dagger}}-\hat{a}_i^\dagger\hat{a}_j^\dagger\hat{a}_b^{\phantom{\dagger}}\hat{a}_a^{\phantom{\dagger}}\big)}|\Psi_0\rangle\nonumber\\
    &~~~~~=\cos\theta_{ij}^{ab}\cos\theta_{il}^{cd}e^{\tan\theta_{il}^{cd}\hat{a}_c^\dagger\hat{a}_d^\dagger\hat{a}_l^{\phantom{\dagger}}\hat{a}_i^{\phantom{\dagger}}+\tan\theta_{ij}^{ab}\sec\theta_{il}^{cd}\hat{a}_a^\dagger\hat{a}_b^\dagger\hat{a}_j^{\phantom{\dagger}}\hat{a}_i^{\phantom{\dagger}}}|\Psi_0\rangle.
\end{align}
One can check that this gives the correct result. Note, that this is in the standard CC form because we could remove the operator exponent on the cosine factor in the amplitude by acting on the reference state before putting all excitation terms in the same exponent. 

We show one final example, consisting of two doubles that commute with each other (so they excite a quad), followed by another double, which has a de-excitation from the quad. The re-ordering of the first two terms is simple, because they commute with each other. We choose the first double to have indices $ij;ab$, the second to be $kl;cd$ and the third to be $ik;ac$; we assume, for simplicity, that $i<j<k<l<a<b<c<d$. As before, for the first two terms, the de-excitation term annihilates against the reference and the exponentials involving the projection operators will evaluate to single powers of cosines as well. Using the factorizations we have already shown, and the fact that no indices in the first two terms are in common, we have that the application of these three terms can be written as
\begin{align}
    &e^{\theta_{ik}^{ac}\big (\hat{a}_a^\dagger\hat{a}_a^\dagger\hat{a}_k^{\phantom{\dagger}}\hat{a}_i^{\phantom{\dagger}}-\hat{a}_i^\dagger\hat{a}_k^\dagger\hat{a}_c^{\phantom{\dagger}}\hat{a}_a^{\phantom{\dagger}}\big)}
    e^{\theta_{kl}^{cd}\big (\hat{a}_c^\dagger\hat{a}_d^\dagger\hat{a}_l^{\phantom{\dagger}}\hat{a}_k^{\phantom{\dagger}}-\hat{a}_k^\dagger\hat{a}_l^\dagger\hat{a}_d^{\phantom{\dagger}}\hat{a}_c^{\phantom{\dagger}}\big)}
    e^{\theta_{ij}^{ab}\big (\hat{a}_a^\dagger\hat{a}_b^\dagger\hat{a}_j^{\phantom{\dagger}}\hat{a}_i^{\phantom{\dagger}}-\hat{a}_i^\dagger\hat{a}_j^\dagger\hat{a}_b^{\phantom{\dagger}}\hat{a}_a^{\phantom{\dagger}}\big)}|\Psi_0\rangle\nonumber\\
    &~~~~~=\cos\theta_{ij}^{ab}\cos\theta_{kl}^{cd}e^{\tan\theta_{ik}^{ac}\hat{a}_a^\dagger\hat{a}_a^\dagger\hat{a}_k^{\phantom{\dagger}}\hat{a}_i^{\phantom{\dagger}}}
    e^{-\ln\big(\cos\theta_{ik}^{ac}\big )\big (\hat{n}_a\hat{n}_c(1-\hat{n}_i)(1-\hat{n}_k)-(1-\hat{n}_a)(1-\hat{n}_c)\hat{n}_i\hat{n}_k\big)}\nonumber\\
    &~~~~~\times
    e^{-\tan\theta_{ik}^{ac}\hat{a}_i^\dagger\hat{a}_k^\dagger\hat{a}_c^{\phantom{\dagger}}\hat{a}_a^{\phantom{\dagger}}}e^{\tan\theta_{kl}^{cd}\hat{a}_c^\dagger\hat{a}_d^\dagger\hat{a}_l^{\phantom{\dagger}}\hat{a}_k^{\phantom{\dagger}}+\tan\theta_{ij}^{ab}\hat{a}_a^\dagger\hat{a}_b^\dagger\hat{a}_j^{\phantom{\dagger}}\hat{a}_i^{\phantom{\dagger}}}|\Psi_0\rangle.
\end{align}
First, we need to move the de-excitation operator through the two excitation operators to the right in the last line. We do this one operator at a time so that the commutator correction terms commute with the other operator in the exponent. This gives
\begin{align}
    &e^{-\tan\theta_{ik}^{ac}\hat{a}_i^\dagger\hat{a}_k^\dagger\hat{a}_c^{\phantom{\dagger}}\hat{a}_a^{\phantom{\dagger}}}e^{\tan\theta_{kl}^{cd}\hat{a}_c^\dagger\hat{a}_d^\dagger\hat{a}_l^{\phantom{\dagger}}\hat{a}_k^{\phantom{\dagger}}+\tan\theta_{ij}^{ab}\hat{a}_a^\dagger\hat{a}_b^\dagger\hat{a}_j^{\phantom{\dagger}}\hat{a}_i^{\phantom{\dagger}}}=e^{\tan\theta_{kl}^{cd}\hat{a}_c^\dagger\hat{a}_d^\dagger\hat{a}_l^{\phantom{\dagger}}\hat{a}_k^{\phantom{\dagger}}}e^{-\tan\theta_{ik}^{ac}\tan\theta_{kl}^{cd}\hat{a}_i^\dagger\hat{a}_d^\dagger\hat{a}_a^{\phantom{\dagger}}\hat{a}_l^{\phantom\dagger}(\hat{n}_c-\hat{n}_k)}\nonumber\\
    &~~~~~\times
    e^{-\tan\theta_{ik}^{ac}\hat{a}_i^\dagger\hat{a}_k^\dagger\hat{a}_c^{\phantom{\dagger}}\hat{a}_a^{\phantom{\dagger}}}e^{\tan\theta_{ij}^{ab}\hat{a}_a^\dagger\hat{a}_b^\dagger\hat{a}_j^{\phantom{\dagger}}\hat{a}_i^{\phantom{\dagger}}}\nonumber\\
    &e^{-\tan\theta_{ik}^{ac}\hat{a}_i^\dagger\hat{a}_k^\dagger\hat{a}_c^{\phantom{\dagger}}\hat{a}_a^{\phantom{\dagger}}}e^{\tan\theta_{kl}^{cd}\hat{a}_c^\dagger\hat{a}_d^\dagger\hat{a}_l^{\phantom{\dagger}}\hat{a}_k^{\phantom{\dagger}}+\tan\theta_{ij}^{ab}\hat{a}_a^\dagger\hat{a}_b^\dagger\hat{a}_j^{\phantom{\dagger}}\hat{a}_i^{\phantom{\dagger}}}=e^{\tan\theta_{kl}^{cd}\hat{a}_c^\dagger\hat{a}_d^\dagger\hat{a}_l^{\phantom{\dagger}}\hat{a}_k^{\phantom{\dagger}}}e^{-\tan\theta_{ik}^{ac}\tan\theta_{kl}^{cd}\hat{a}_i^\dagger\hat{a}_d^\dagger\hat{a}_a^{\phantom{\dagger}}\hat{a}_l^{\phantom\dagger}(\hat{n}_c-\hat{n}_k)}\nonumber\\
    &~~~~~\times
    e^{\tan\theta_{ij}^{ab}\hat{a}_a^\dagger\hat{a}_b^\dagger\hat{a}_j^{\phantom{\dagger}}\hat{a}_i^{\phantom{\dagger}}}
    e^{-\tan\theta_{ik}^{ac}\tan\theta_{ij}^{ab}\hat{a}_k^\dagger\hat{a}_b^\dagger\hat{a}_d^{\phantom{\dagger}}\hat{a}_j^{\phantom\dagger}(\hat{n}_a-\hat{n}_i)}
    e^{-\tan\theta_{ik}^{ac}\hat{a}_i^\dagger\hat{a}_k^\dagger\hat{a}_c^{\phantom{\dagger}}\hat{a}_a^{\phantom{\dagger}}}\nonumber\\
    &e^{-\tan\theta_{ik}^{ac}\hat{a}_i^\dagger\hat{a}_k^\dagger\hat{a}_c^{\phantom{\dagger}}\hat{a}_a^{\phantom{\dagger}}}e^{\tan\theta_{kl}^{cd}\hat{a}_c^\dagger\hat{a}_d^\dagger\hat{a}_l^{\phantom{\dagger}}\hat{a}_k^{\phantom{\dagger}}+\tan\theta_{ij}^{ab}\hat{a}_a^\dagger\hat{a}_b^\dagger\hat{a}_j^{\phantom{\dagger}}\hat{a}_i^{\phantom{\dagger}}}=e^{\tan\theta_{kl}^{cd}\hat{a}_c^\dagger\hat{a}_d^\dagger\hat{a}_l^{\phantom{\dagger}}\hat{a}_k^{\phantom{\dagger}}}e^{\tan\theta_{ij}^{ab}\hat{a}_a^\dagger\hat{a}_b^\dagger\hat{a}_j^{\phantom{\dagger}}\hat{a}_i^{\phantom{\dagger}}}\nonumber\\
    &~~~~~\times
    e^{-\tan\theta_{ik}^{ac}\tan\theta_{kl}^{cd}\hat{a}_i^\dagger\hat{a}_d^\dagger\hat{a}_a^{\phantom{\dagger}}\hat{a}_l^{\phantom\dagger}(\hat{n}_c-\hat{n}_k)}
    e^{-\tan\theta_{ik}^{ac}\tan\theta_{kl}^{cd}\tan\theta_{ij}^{ab}\hat{a}_b^\dagger\hat{a}_d^\dagger\hat{a}_l^{\phantom{\dagger}}\hat{a}_j^{\phantom\dagger}(\hat{n}_a-\hat{n}_i)(\hat{n}_c-\hat{n}_k)}\nonumber\\
    &~~~~~\times
    \cancel{e^{-\tan\theta_{ik}^{ac}\tan\theta_{ij}^{ab}\hat{a}_k^\dagger\hat{a}_b^\dagger\hat{a}_d^{\phantom{\dagger}}\hat{a}_j^{\phantom\dagger}(\hat{n}_a-\hat{n}_i)}}
    \cancel{e^{-\tan\theta_{ik}^{ac}\hat{a}_i^\dagger\hat{a}_k^\dagger\hat{a}_c^{\phantom{\dagger}}\hat{a}_a^{\phantom{\dagger}}}},
\end{align}
because we need only the single commutator correction term. Here, we cancelled the last two terms, which vanish when they act on the reference state. There is a new excitation operator, multiplied by projection operators as the second term in the next to last line. Because the projection operators commute with the rest of the operator, and because they evaluate to one when they act on the reference state, they can be removed from that term. Then, the leftmost term on the second to last line commutes with the term to the right, and after re-ordering, it is removed when it operates on the reference state. So, after acting on the reference, we are left with three pure excitation terms, given by
\begin{equation}
    e^{\tan\theta_{kl}^{cd}\hat{a}_c^\dagger\hat{a}_d^\dagger\hat{a}_l^{\phantom{\dagger}}\hat{a}_k^{\phantom{\dagger}}}e^{\tan\theta_{ij}^{ab}\hat{a}_a^\dagger\hat{a}_b^\dagger\hat{a}_j^{\phantom{\dagger}}\hat{a}_i^{\phantom{\dagger}}}e^{-\tan\theta_{ik}^{ac}\tan\theta_{kl}^{cd}\tan\theta_{ij}^{ab}\hat{a}_b^\dagger\hat{a}_d^\dagger\hat{a}_l^{\phantom{\dagger}}\hat{a}_j^{\phantom\dagger}}.
\end{equation}
Each one of these terms must be re-ordered against the projection term; the third factor already commutes with that term and can just be re-ordered. As we have seen before, the projection term renormalizes the coefficient of the excitation term and the projection term evaluates to a cosine against the reference state. Following a similar calculation as before, we find that our final result becomes
\begin{align}
    &e^{\theta_{ik}^{ac}\big (\hat{a}_a^\dagger\hat{a}_a^\dagger\hat{a}_k^{\phantom{\dagger}}\hat{a}_i^{\phantom{\dagger}}-\hat{a}_i^\dagger\hat{a}_k^\dagger\hat{a}_c^{\phantom{\dagger}}\hat{a}_a^{\phantom{\dagger}}\big)}
    e^{\theta_{kl}^{cd}\big (\hat{a}_c^\dagger\hat{a}_d^\dagger\hat{a}_l^{\phantom{\dagger}}\hat{a}_k^{\phantom{\dagger}}-\hat{a}_k^\dagger\hat{a}_l^\dagger\hat{a}_d^{\phantom{\dagger}}\hat{a}_c^{\phantom{\dagger}}\big)}
    e^{\theta_{ij}^{ab}\big (\hat{a}_a^\dagger\hat{a}_b^\dagger\hat{a}_j^{\phantom{\dagger}}\hat{a}_i^{\phantom{\dagger}}-\hat{a}_i^\dagger\hat{a}_j^\dagger\hat{a}_b^{\phantom{\dagger}}\hat{a}_a^{\phantom{\dagger}}\big)}|\Psi_0\rangle\nonumber\\
    &~~~~~=\cos\theta_{ij}^{ab}\cos\theta_{kl}^{cd}\cos\theta_{ik}^{ac}e^{\tan\theta_{ik}^{ac}\hat{a}_a^\dagger\hat{a}_a^\dagger\hat{a}_k^{\phantom{\dagger}}\hat{a}_i^{\phantom{\dagger}}}e^{\tan\theta_{kl}^{cd}\big(\cos\theta_{ik}^{ac}\big)^{-\hat{n}_a(1-\hat{n}_i)-(1-\hat{n}_a)\hat{n}_i}\hat{a}_c^\dagger\hat{a}_d^\dagger\hat{a}_l^{\phantom{\dagger}}\hat{a}_k^{\phantom{\dagger}}}\nonumber\\
    &~~~~~\times e^{\tan\theta_{ij}^{ab}\sec\theta_{ik}^{ac}\hat{a}_a^\dagger\hat{a}_b^\dagger\hat{a}_j^{\phantom{\dagger}}\hat{a}_i^{\phantom{\dagger}}}
    e^{-\tan\theta_{ik}^{ac}\tan\theta_{kl}^{cd}\tan\theta_{ij}^{ab}\hat{a}_b^\dagger\hat{a}_d^\dagger\hat{a}_l^{\phantom{\dagger}}\hat{a}_j^{\phantom\dagger}}|\Psi_0\rangle.
\end{align}
This final result is important, because it shows that one cannot immediately remove the operator term in the exponent of the amplitude of the excitation operator. This is because it does not commute with the excitation term to the right of it. It can be removed by picking the proper value of the exponent when it acts directly on the reference, and adding in higher-rank terms to fix coefficients of determinants with the wrong exponent, but we will not show further details for how this can be done in this work, as it is not critical that we show that the final form of the CC ansatz can be written in the traditional CC form.

%%%%%%%%%%%%%%%%%%%%%%%%%%%%%%%%%%%%%%%%%%
\section{Discussion}

We now discuss why this process should work, even though we have not shown how it works in the most general case. The reason is rather simple. We can take the factorized form of the UCC and use the Euler-identity like relation, from Eq.~(\ref{eq:euler}), which breaks each UCC factor into either an identity operator, or a term that has a cosine multiplying one term and a sine multiplying another. This means when we have created an intermediate ansatz wavefunction, by acting a number of UCC factors onto the reference, we find that the next UCC factor will either leave a determinant in the superposition unchanged, or will multiply it by a cosine and add a second determinant that is multiplied by a sine. This means all terms in the expansion for the wavefunction involve polynomials in sines and cosines of the different angles. As we saw in our examples, we expect the transformation from UCC to CC to have a prefactor of a cosine of the angle for each UCC factor. This will change many of the cosine factors in the polynomial coefficients to ones and will change sines to tangents (it may also introduce some secant terms).  But there are no other types of terms that can be created. Furthermore, all of the determinants in the UCC ansatz are excitations from the single reference. This implies that either, we never run into any of the ``edge'' cases where separating out the different exponents after reordering two operators is one of the complicated cases, or the separation can be performed, yielding a simple final result. While not a rigorous proof, this is a strong indication that the above algorithm will always work to convert the UCC ansatz to the CC ansatz, even if edge cases do occur. But note, this argument does not preclude that the CC form may need to be a factorized form and that the amplitudes may be raised to operator-valued integer powers. 

Another point worth discussing is about the situation when we approximate the traditional UCC by a Trotter formula with some number of Trotter steps. In this case, we will eventually need to move excitation operators past their precise de-excitation counterpart, because of the repeating nature of the factors in the Trotterized form. This is one of the edge cases we have been discussing ($\hat{A}$ matches $\hat{A}'$ and $\hat{A}'$ matches $\hat{A}$), but this case can be handled easily, because it is well known that one can create disentangling identities in any order for the three factors (exponentials of $\hat{S}_+$, $\hat{S}_z$, and $S_-$)~\cite{disentangling-ordering}. If we have to reorder Trotter factors, we would be presented with a product of the form $z$, $-$, $+$. We simply use the different identities to turnover the product into $+$, $z$, $-$, which is guaranteed to be possible. So, such a situation will not cause any problems for carrying out the conversion to the factorized CC form either.

Another point we have found in our analysis is that the conversion from UCC to CC often requires additional terms than those that were in the original UCC ansatz. It is possible that these new terms could be the same rank as the terms that were being re-ordered (if they have a definite rank), or it can be a higher rank), although creating a higher-rank term that does not annihilate against the reference will require terms arising from many re-orderings. In addition, many of the terms we create as we re-order will annihilate against the reference, this keeps the identity (which is an exact operator identity) from becoming too large, which it easily can if we do not apply to the reference. We also expect that a UCC ansatz that is low rank, involving excitation-de-excitation operators all less than some specific rank, will likely map to a CC ansatz that has many higher-rank terms. In fact, it is likely that the number of amplitudes that we find in the mapped CC ansatz are as many as the determinants in the ansatz wavefunction, because the exponential form that arises from the CC ansatz does not produce the same amplitudes for higher-rank determinants as the UCC does---then an additional higher-rank amplitude is needed to produce the same ansatz wavefunction. Finally, because some of the amplitudes determined in this operator identity depend on the determinants that they operate on, we have to use a factorized form for the CC ansatz, because not all factors commute with each other. This is able to be fixed by employing additional higher rank factors, but we do not pursue that more thoroughly here.

The fact that the UCC approximation is unitary (hence the state remains unit norm) and the CC approximation can be extracted from the UCC in the way we describe, with an additional product of cosine factors, suggests that the norm of the CC state is larger than one (since a product of cosines is always less than one). It is not clear there is anything physically meaningful in this statement, but it is an observation that can be made. Finally, the UCC approximation is variational and it maps to a similar CC approximation, but including many terms of higher rank. What this suggests, however, is that there may be a way to restore the variational nature of the approximation to a CC ansatz by modifying the ansatz to not be low rank. It is not clear there is any simple way to determine how to do the modification, or if it requires adding in too many additional amplitudes to be practical, but it is an interesting insight brought forward by this analysis.

The Euler-identity like relation for the factorized form of the UCC suggest an alternative way to relate to the CC. We first compute the desired ansatz wavefunction by using the Euler-identity like relation. This creates the full wavefunction ansatz. We then apply an algorithm similar in spirit to the elimination algorithm~\cite{evangalista}, but able to be applied much more easily for the CC situation. One starts from the determinants that can be reached by applying a rank-one operator to the reference. We pick the amplitude of the determinant to be the amplitude of the CC term. Then, we compute the exponential of the CC operator and subtract off all of the amplitudes for any higher-rank determinants that are generated by applying the rank-one CC operator to the reference. Then we repeat with the rank-two amplitudes and so forth. Because we do not expect the higher-rank terms to be created by a low-rank CC operator, this procedure will continue up to the highest-rank determinant in the UCC ansatz wavefunction. This is, in part, the reason why the manipulations for the re-ordering create so many new operators.

An important question is whether the additional terms arising from the re-ordering are important or can be neglected. If the amplitudes are all small angles, then these higher-order terms will typically include higher-powers of tangents or sines of these angles. If the angles are small, these powers can become very small quickly---hence, there is a possibility for weakly correlated molecules, that the correction terms are not important. Only quantitative analysis and a clear error bound can resolve how important the extra terms are. But, because we expect at least some of the amplitudes to be large, we do anticipate at least some of the correction terms to be large enough to be important. Working out examples of that is beyond what will be done in this paper.

An alternative approach that one can try is to form the commutator algebra by commuting all of the operators of a given class that is included in the UCC ansatz to determine the closed commutator algebra of the system. If there are a total of $N$ spin orbitals, then this operator algebra is a subalgebra of the SU($2^N$) Lie algebra. One can then perform a Cartan decomposition of the Lie algebra (assuming one finds an appropriate involution) and use the KHK construction to compute the UCC ansatz as well~\cite{izmaylov}. Then one can try to convert the KHK form into a CC form. But it is unlikely that one could work out such a program, even for a small system, because the Lie groups become too large too quickly (once doubles or higher-rank operators are included, the commutator algebra tends to become closed only for the full Lie algebra).

So, the best way to carry out this algorithm is on a computer following the rules we developed above. Because the UCC ansatz in the factorized form can involve an exponentially large number of terms, again the algorithm is likely to be limited to smaller-size systems as well. And how do we find the connection between UCC and CC for the traditional UCC? We simply use the Trotter formula with a large enough number of steps to accurately produce the conventional form of the UCC. So, while not providing an analytical formula connecting the amplitudes of the two ansatzes, the approach outlined in this work helps us understand how the two ansatzes (CC versus UCC) compare to each other.

%%%%%%%%%%%%%%%%%%%%%%%%%%%%%%%%%%%%%%%%%%
\section{Conclusions}

In this work, we have shown how one can relate a single-reference UCC ansatz in factorized form to its corresponding single-reference CC ansatz in a factorized form. By using the Trotter product formula, this approach can be extended to also include the traditional UCC ansatz, and by adding additional higher-rank terms, the factorized form of the CC ansatz can be converted to the traditional form. This resolves a longstanding issue in quantum chemistry about how the UCC and CC ansatzes relate. But, determining an explicit formula for a particular ansatz is too complicated to be carried out analytically. It is a straightforward exercise to perform the algebraic manipulations using a computer and this would make an interesting follow-up study for systems where the wavefunction does not have too many determinants contributing to it. 

This work is important for quantum computing because it shows how a common quantum computing ansatz relates to the conventional computing ansatz based on the CC approach. This allows for better understanding of the accuracy we anticipate we will be able to achieve with quantum computing.

%%%%%%%%%%%%%%%%%%%%%%%%%%%%%%%%%%%%%%%%%%
\vspace{6pt} 

%%%%%%%%%%%%%%%%%%%%%%%%%%%%%%%%%%%%%%%%%%
%% optional
%\supplementary{The following are available online at \linksupplementary{s1}, Figure S1: title, Table S1: title, Video S1: title.}


%%%%%%%%%%%%%%%%%%%%%%%%%%%%%%%%%%%%%%%%%%

\funding{This research was funded by the National Science Foundation under Grant No. CHE-1836497 and  the McDevitt bequest at Georgetown University. This research was also supported in part by the National Science Foundation under Grant No. NSF PHY-1748958 in the context of a KITP program ``Towards Classically Intractable Quantum Simulations of Physics and Chemistry.''}

\dataavailability{There is no data used in this study.} 

\acknowledgments{I acknowledge useful discussions with Rod Bartlett, Jia Chen, Hai-Ping Cheng, Avijit Shen, and Dominika Zgid.}

\conflictsofinterest{The author declares no conflict of interest.}

%%%%%%%%%%%%%%%%%%%%%%%%%%%%%%%%%%%%%%%%%%
\end{paracol}
\reftitle{References}


\begin{thebibliography}{99}
\bibitem{bartlett_review}
Bartlett, R. J. and Musia\l, M. Coupled-cluster theory in quantum chemistry.
\textit{Rev. Mod. Phys.} \textbf{2007}, \textit{79}, 291--352.
\bibitem{kutzelnig}
Kutzelnigg, W. Pair Correlation Theories, In \textit{Methods of Electronic Structure Theory}; Schaefer III,
H. F., Ed.; Springer US: New York, NY, USA, 1977; Chapter 5, pp.
129--188.
\bibitem{bartlett}
Bartlett, R. J.; Kucharski, S. A.; Noga, J. Alternative coupled-cluster ansatze II. The unitary coupled-cluster method. \textit{Chem. Phys.
Lett.} \textbf{1989}, \textit{155}, 133--140.
\bibitem{paldus}
Li, X.; Paldus, J. Unitary Group Based Coupled Cluster Methods and Calculation of Molecular Properties. In \textit{Recent Advances in Coupled-Cluster Methods}; Bartlett, R. J., Ed. Series \textit{Recent Advances in Computational Chemistry}; \textit{Vol. 3}; World Scientific: Singapore, 1997; pp. 183-219.
\bibitem{scuseria}
Harsha, G.;  Shiozaki, T.; Scuseria, G. E.
On the difference between variational and unitary coupled cluster theories.
\textit{J. Chem. Phys.} \textbf{2018}, \textit{148}, 044107. 
\bibitem{mazziotti}
Mazziotti, D. A.
Anti-Hermitian Contracted Schr\"odinger Equation: Direct Determination of the Two-Electron Reduced Density Matrices of Many-Electron Molecules.
\textit{Phys. Rev. Lett.}  \textbf{2006}, \textit{97}, 143002.
\bibitem{vqe}
 Peruzzo, A.; McClean, J.; Shadbolt, P.; Yung, M.-H.; Zhou, X.-Q.; Love, P. J.; Aspuru-Guzik, A.; O’Brien, J. L. 
 A variational eigenvalue solver on a photonic quantum processor. \textit{Nat. Commun.}
\textbf{2014}, \textit{5}, 4213.
\bibitem{adapt}
Grimsley, H. R.; Economou, S. E.; Barnes, E.; Mayhall, N. J. An
adaptive variational algorithm for exact molecular simulations on a
quantum computer. \textit{Nat. Commun.} \textbf{2019}, \textit{10}, 3007.
\bibitem{evangalista}
 Evangelista, F. A.; Chan, G. K.-L.; Scuseria, G. E. Exact
parameterization of fermionic wave functions via unitary coupled
cluster theory. \textit{J. Chem. Phys.} \textbf{2019}, \textit{151}, 244112.
\bibitem{xu}
Xu, L.; Lee, J. T.; Freericks, J. K. Test of the unitary coupled-cluster variational quantum eigensolver for a simple strongly
correlated condensed-matter system. \textit{Mod. Phys. Lett. B} \textbf{2020}, \textit{34},
2040049.
\bibitem{jia}
Chen, J.; Cheng, H.-P.; Freericks, J. K.  Quantum-Inspired Algorithm for the Factorized Form of Unitary Coupled Cluster Theory, \textit{J. Chem. Theor. Comp.} \textbf{2021}, \textit{17}, 841--847.
\bibitem{atom}
Arecchi, F. T.; Courtens, E.; Gilmore, R.; Thomas, H. Atomic Coherent States in Quantum Optics. \textit{Phys. Rev. A} \textbf{1972}, \textit{6}, 2211--2237.
\bibitem{ucc_exp_sum}
Cooper, B.; Knowles, P. J.
Benchmark studies of variational, unitary and extended coupled cluster methods.
\textit{J. Chem. Phys.} \textbf{2010}, \textit{133}, 234102. 
\bibitem{ucc3}
Chen, J.; Cheng, H.-P.; Freericks, J. K.  Flexibility of the factorized form of the unitary coupled cluster ansatz. \textit{J. Chem. Phys.} \textbf{2022}, \textit{156}, 044106.
\bibitem{jacobsen}
Jacobson, N. \textit{Lie Algebras}. Interscience Publishers: New York, NY, USA, 1962.
\bibitem{pascual}
Galindo, A. and Pascual, P. \textit{Quantum Mechanics I}. Springer-Verlag, Inc., Berlin, Germany, 1990.
\bibitem{disentangling-ordering}
Mufti, A.; Schmitt, H. A.;  and Sargent III, M. Finite-dimensional matrix representations as
calculational tools in quantum optics, \textit{Am. J. Phys.} \textbf{1991}, \textit{61}, 729--733.
\bibitem{izmaylov}
Izmaylov, A. F.; Manuel Díaz-Tinoco, M.; Lang, R. A.
On the order problem in construction of unitary operators for the variational quantum eigensolver	\textit{Phys. Chem. Chem. Phys.} \textbf{2020}, \textit{22}, 12980--12986.


\end{thebibliography}
\end{document}